\documentclass[apj]{emulateapj}
\usepackage{graphicx}
\usepackage{amsmath}
\usepackage[usenames]{color}
\graphicspath{{../figures/}}

\newcommand{\om}{\Omega_m}
\newcommand{\ok}{\Omega_k}

\newcommand{\orc}{\Omega_{r_c}}

\newcommand{\beq}{\begin{equation}}
\newcommand{\eeq}{\end{equation}}
\newcommand{\bea}{\begin{eqnarray}}
\newcommand{\eea}{\end{eqnarray}}

\shorttitle{Using SDSS-II Supernovae to Test Exotic Cosmologies}
\shortauthors{Sollerman, M\"ortsell, Davis, Blomqvist, et al.}

\begin{document}

\title{First-Year Sloan Digital Sky Survey-II (SDSS-II)
Supernova Results: 
Constraints on Non-Standard Cosmological Models}

\author{
{J.~Sollerman}\altaffilmark{1,2},
{E.~M\"ortsell}\altaffilmark{3},
{T.~M.~Davis}\altaffilmark{1,4},
{M. Blomqvist}\altaffilmark{2},
{B.~Bassett}\altaffilmark{5,6}, 
{A.~C.~Becker}\altaffilmark{7},
{D.~Cinabro}\altaffilmark{8},       
{A.~V.~Filippenko}\altaffilmark{9},
{R.~J.~Foley}\altaffilmark{9,10,11},       
{J.~Frieman}\altaffilmark{12,13,14},    
{P.~Garnavich}\altaffilmark{15},
{H.~Lampeitl}\altaffilmark{16},
{J.~Marriner}\altaffilmark{14},
{R.~Miquel}\altaffilmark{17,18},
{R.~C.~Nichol}\altaffilmark{16},
{M.~W.~Richmond}\altaffilmark{19},   
{M.~Sako}\altaffilmark{20},   
{D.~P.~Schneider}\altaffilmark{21},
{M.~Smith}\altaffilmark{5,6,16},
{J.~T.~Vanderplas}\altaffilmark{7}, and
{J.~C.~Wheeler}\altaffilmark{22}
}
\altaffiltext{1}{Dark Cosmology Centre, Niels Bohr Institute, University of Copenhagen, Juliane Maries Vej 30, Copenhagen, Denmark.}
\altaffiltext{2}{The Oskar Klein Centre, Department of Astronomy, AlbaNova, Stockholm University, 106 91 Stockholm, Sweden.}
\altaffiltext{3}{The Oskar Klein Centre, Department of Physics, AlbaNova, Stockholm University, 106 91 Stockholm, Sweden.}
\altaffiltext{4}{School of Mathematics and Physics, University of Queensland, QLD, 4072, Australia.}
\altaffiltext{5}{South African Astronomical Observatory, P.O. Box 9, Observatory 7935, South Africa.}
\altaffiltext{6}{Department of Mathematics and Applied Mathematics, University of Cape Town, Rondebosch 7701, South Africa.}
\altaffiltext{7}{Department of Astronomy, University of Washington, Box 351580, Seattle, WA 98195.}
\altaffiltext{8}{Department of Physics and Astronomy, Wayne State University, Detroit, MI 48202.}
\altaffiltext{9}{Department of Astronomy, University of California, Berkeley, CA 94720-3411.}
\altaffiltext{10}{Harvard-Smithsonian Center for Astrophysics, 60 Garden Street, Cambridge, MA 02138.}
\altaffiltext{11}{Clay Fellow.}
\altaffiltext{12}{Kavli Institute for Cosmological Physics, The University of Chicago, 5640 South Ellis Avenue Chicago, IL 60637.}
\altaffiltext{13}{Department of Astronomy and Astrophysics, The University of Chicago, 5640 South Ellis Avenue Chicago, IL 60637.}
\altaffiltext{14}{Center for Astrophysics, Fermi National Accelerator Laboratory, P. O. Box 500, Batavia IL 60510.}
\altaffiltext{15}{University of Notre Dame, 225 Nieuwland Science, Notre Dame, IN 46556-5670.}
\altaffiltext{16}{Institute of Cosmology and Gravitation, University Portsmouth, Portsmouth, PO1 3FX, UK.}
\altaffiltext{17}{Instituci\'o Catalana de Recerca i Estudis Avan\c{c}ats, Barcelona, Spain.}
\altaffiltext{18}{Institut de F\'{\i}sica d'Altes Energies, E-08193 Bellaterra, Barcelona, Spain.}
\altaffiltext{19}{Physics Department, Rochester Institute of Technology, Rochester, NY 14623.}
\altaffiltext{20}{Department of Physics and Astronomy, University of Pennsylvania, 209 South 33rd Street, Philadelphia, PA 19104.}
\altaffiltext{21}{Department of Astronomy and Astrophysics, 525 Davey Laboratory, Pennsylvania State University, University Park, PA 16802.}      
\altaffiltext{22}{Department of Astronomy, University of Texas, Austin, TX 78712.}

\begin{abstract}
We use the new Type Ia supernovae (SNe~Ia) discovered by
the Sloan Digital Sky Survey-II Supernova Survey together
with additional supernova datasets as well as observations of
the cosmic microwave background and baryon acoustic oscillations
to constrain cosmological models. 
This complements the standard cosmology 
analysis presented by Kessler et al.~(2009)
in that we discuss and rank a number of the most popular non-standard cosmology
scenarios.
When this combined dataset is analyzed using the MLCS2k2 light-curve 
fitter, we find that more exotic models for cosmic acceleration 
provide a better fit to the data than the $\Lambda$CDM model. For example, 
the flat 
Dvali-Gabadadze-Porrati model is ranked 
higher by our information-criteria tests
than the standard model with a flat universe and a cosmological constant.
When the supernova dataset is instead analyzed using the SALT-II 
light-curve fitter, the standard cosmological-constant model fares best.
This investigation of how sensitive 
cosmological model selection is to 
assumptions about, and within, the light-curve fitters 
thereby highlights the need for an improved
understanding of these unresolved systematic effects. 
Our investigation also includes inhomogeneous 
Lema\^{i}tre-Tolman-Bondi (LTB) models.
While our LTB models can be made to fit the supernova 
data as well as any other model, 
the extra parameters they require are not supported by our 
information-criteria analysis.
Finally, we explore more model-independent ways to investigate the cosmic 
expansion based on this new dataset.
\end{abstract}


\section{Introduction}\label{intro}

The Type Ia supernova (SN~Ia) measurements which first indicated an
accelerating expansion of the universe 
(\citealt{riess98}; \citealt{perlmutter99}; 
see \citealt{filippenko05} for a review)
have been confirmed and substantiated by a second generation of
high-redshift supernova experiments \citep{riess04,astier06,WV07,riess07}.  
An important step forward in this respect was recently taken with the
Sloan Digital Sky Survey-II (SDSS-II) 
Supernova Survey \citep[][]{frieman08}.  
This three-year survey, undertaken with a large CCD camera on a 
dedicated 2.5-m telescope in New
Mexico \citep{gunn06}, has discovered and followed several 
hundred SNe~Ia, mainly in the redshift interval 
$z$ = [0.01, 0.45]. These intermediate
redshifts were previously underexplored, and filling this ``redshift
desert'' not only provides important new constraints on cosmology
\citep{kessler09}, but will also help constrain systematic effects by
bridging the low-$z$ and the high-$z$ supernova populations.

The first-year 
SDSS supernova dataset is discussed in three companion papers, 
including this one.
\citet[][ hereafter K09]{kessler09} present the dataset in detail 
and also use it to constrain standard cosmological models.
\cite{hubert09} combine the SDSS SN data with
other constraints
to derive joint constraints on dark energy
from low-redshift ($z<0.4$) measurements only; they also explore
the consistency of the SN and BAO distance scales.

To complement and extend the cosmological analysis presented in these 
papers we will in this paper use the first-year SDSS-II supernova data
to explore several alternative cosmological models. Following the analysis
outlined by \citet[][ hereafter D07]{davis07}, we combine the 103
new SDSS-II SNe~Ia in the K09 dataset with new analyses of 
previously available SN~Ia datasets, 
as well as complementary data, to explore non-standard
cosmologies.  We also investigate the use of more model-free
approaches in constraining the evolution of the universe.

This paper is organized as follows. In Section~2 we describe the 
datasets invoked in the analysis and how they are combined, 
while in Section~3 we present the cosmological 
models that are investigated in this work.
In Section~4 we discuss our results. Section~5 includes a discussion of 
systematic effects, 
while Section~6 presents some ways of expressing generalized
parameters from the supernova dataset.
Finally, in Section~7 we
provide a summary of our results. 

When we refer to the ``standard model'' we mean
Friedmann-Robertson-Walker cosmology with a constant dark energy
equation-of-state parameter, also known as ``$w$ Cold Dark Matter''
($w$CDM), of which the cosmological-constant model ($\Lambda$CDM) is a
special case.  We use units in which $c=1$.

\section{Datasets}\label{datasets}

In this paper we make use of constraints from several different datasets
in order to test a number of cosmological models. 
Compared to the previous analysis by D07, here we make use of an 
enlarged and re-analyzed supernova set, a new prescription as well as 
new data for the baryonic acoustic oscillations (BAO), and also an updated 
cosmic microwave background (CMB) analysis.
In this section we present these datasets, and describe how they are combined.

\subsection{Type Ia Supernovae}

The primary new dataset used in this analysis comprises the 103 new SNe~Ia
from the first-year SDSS-II supernova survey 
(\citealt{frieman08}; \citealt{sako08}; K09).       

This sample is published and discussed at length in our companion paper 
(K09) which
also includes a comprehensive and consistent re-analysis of 
other sets of local and high-$z$ SNe~Ia, using the same light-curve fitter.
This is important since it 
ensures that all the supernova datasets are
treated in a uniform manner regarding selection criteria and 
light-curve fitting. For the analysis presented in this paper 
we start by discussing the supernova dataset analyzed using
the Multicolor Light Curve Shape 2k2 fitter: 
(MLCS; Jha et al. 2007)\footnote{We have used the entire Nearby+SDSS+ESSENCE+SNLS+HST dataset (e) of K09; http://das.sdss.org/va/SNcosmology/sncosm09$\_$fits.tar.gz.}.
According to the MLCS analysis of K09, the
standard $\Lambda$CDM cosmological model provides a 
rather poor fit to these data.
We will also discuss
calculations for the K09 SN dataset analyzed with the SALT-II light-curve
fitter \citep{guy}, which is better fit by the $\Lambda$CDM concordance model.
In total, we use 288 SNe from the analysis of K09. The 
distance moduli and redshifts for these SNe
are provided in K09 
(their Table~10 for MLCS and Table~14 for SALT-II).
The very detailed and restrictive 
selection criteria for the total sample
of SNe are described in K09; following their analysis
we have used only supernovae with $z>0.02$ and added an additional
``intrinsic'' dispersion of $\sigma_{\rm{add}}=0.16$ mag to the 
uncertainties output by the light-curve fitter.
 Since this added dispersion is motivated in K09 to make the 
{\it local} MLCS SN Hubble diagram have 
a $\chi^{2}$ equal to unity per degree of freedom, it is essentially
independent of cosmology.
A similar intrinsic dispersion (we use 0.14~mag) gives 
a $\chi^{2}_{\rm d.o.f.}$=1 for the global SALT-II fit.
Our supernova dataset is thus a well-selected sample compiled from many surveys
\citep[][ and references therein]{kessler09,WV07,astier06,riess07,jha07}.

\subsection{Cosmic Microwave Background}
When analyzing CMB observations there are two useful 
parameters commonly employed.  
One describes the scaled distance to recombination, ${\cal R}$,
and the other the angular scale of the sound horizon at
recombination, $\ell_A$ \citep[e.g.,][]{komatsu08,elgaroy07,wang06}. 

The shift parameter, ${\cal R}$, is given by
\beq 
{\cal R} = \sqrt{\frac{\om}{|\ok|}} 
S_k\left[ \sqrt{|\ok|}\int_0^{z_*}\frac{H_0~dz}{H(z)}\right],
\eeq
where $S_k(x) = \sin x, x, \sinh x$ for $k=1,0,-1$, respectively, 
and $z_*$ is the redshift of the last-scattering surface.

The position of the first CMB power-spectrum peak, which represents
the angular scale of the sound horizon at recombination, is given by
\beq \ell_A = \pi \frac{d_{\rm A}(z_*)}{r_{\rm s}(z_*)},\eeq
where $d_{\rm A}(z_*)$ is the comoving angular-diameter distance to 
recombination while
the comoving sound horizon at photon decoupling, $r_{\rm s}$, is given by
\beq r_{\rm s} = \int_{z_*}^\infty \frac{c_s}{H(z)}dz,\label{eq:rs}\eeq 
which depends upon the speed of sound before recombination, $c_s$.
Using both these parameters
in combination reproduces closely the fit from the full CMB power spectrum
\citep[but see also][ for some caveats]{elgaroy07}, 
and within the standard model the two parameters are 
only weakly correlated.
Here we use the recent CMB measurements from the 
five-year Wilkinson Microwave Anisotropy Probe (WMAP) observations, and 
adopt the values $\ell_A=302.10\pm0.86$ and 
${\cal R}=1.710 \pm 0.019$ 
with correlation coefficient $0.1109$ from \citet{komatsu08}.
We further assume $z_*=1090$ exactly 
(Komatsu et al. 2009; variations within the uncertainties 
about this value do not give significant differences in the results). 

In a previous paper (D07), we used only the CMB-${\cal R}$ parameter 
to constrain cosmologies. The same method is used in the K09 paper.
Here we will, for reasons outlined below, adopt a different approach.

Although ${\cal R}$ has been commonly used to constrain
non-standard models, this approach may not always be entirely 
appropriate, because parameters close to standard $w$CDM were assumed 
in deriving the value of ${\cal R}$ 
\citep[see, e.g., Section~5.4.1 of][ Section~6]{komatsu08,kowalski08}. 
We will therefore use ${\cal R}$ {\em only} for the $w$CDM model for
which it was derived. The resulting fit is of interest
because it 
can reveal any tension
between the BAO/CMB and the 
supernova constraints.
As will be further discussed below, $\Lambda$CDM is not 
a good fit to the
data when MLCS is used for the SN analysis, 
and this motivates our search for better fits among non-standard
models.  

We therefore instead perform an analysis without ${\cal R}$ 
for all the models. In doing this, we use only the product of the acoustic
scale $\ell_A$ with the position of the BAO peak (Section~\ref{sect:bao}) 
to complement the SN data.

\subsection{Baryon Acoustic Oscillations}\label{sect:bao}

As with the CMB, there are several parameters in common use for
comparing BAO observations to theoretical models. The most immediately
observable of these is a measurement of the ratio of the sound horizon
scale at the drag epoch, $r_s(z_d)$,
to the dilation scale, $D_V(z)$.
The drag epoch, $z_d\approx1020$, 
is the epoch at which the acoustic oscillations are frozen in.

A more model-independent constraint can be achieved by multiplying the BAO measurement of $r_s(z_d)/D_V(z)$ with the CMB measurement  $\ell_A=\pi
d_A(z_*)/r_s(z_*)$, thus cancelling some of the dependence on the sound horizon scale.

\citet{percival09} measured $r_s(z_d)/D_V(z)$ at two redshifts,
$z=0.2$ and $z=0.35$, finding
$r_s(z_d)/D_V(0.2)=0.1905\pm0.0061$ and
$r_s(z_d)/D_V(0.35)=0.1097\pm0.0036$.
Combining this with $\ell_A$  gives the
combined BAO/CMB-$\ell_A$ constraints:

\bea   
\frac{d_A(z_*)}{D_V(0.2)}\; \frac{r_s(z_d)}{r_s(z_*)}&=& 18.32\pm0.59,\label{eq:dADV_nodrag}\\
\frac{d_A(z_*)}{D_V(0.35)}\; \frac{r_s(z_d)}{r_s(z_*)}  &=& 10.55\pm0.35. \nonumber
\eea

This combination is equivalent to the $S_k/D_V$ combination used in 
\citet{percival07},
but with the ratio of the sound horizon at the two epochs made explicit.
Before matching to cosmological models we also need to implement the correction
for the difference between the sound horizon at the end of the drag
epoch, $z_d\approx1020$, and
the sound horizon at last-scattering, $z_*\approx1090$.  The first is
relevant for the BAO, the second for the CMB, 
and $r_s(z_d)/r_s(z_*)=1.044\pm0.019$ \citep[using values from][]{komatsu08}.
Inserting this into Equation~\ref{eq:dADV_nodrag} gives the final constraints 
we use for the cosmology analysis:
\bea
\frac{d_A(z_*)}{D_V(0.2)} &=&  17.55\pm0.65,\label{eq:dADV_drag}\\
\frac{d_A(z_*)}{D_V(0.35)} &=& 10.10\pm0.38. \nonumber
\eea
We take into account the correlation between these measurements using the
correlation coefficient of 0.337 calculated by \citet{percival09}.

The ratio of sound horizon distances
between drag epoch and last-scattering was  calculated using the
F$\Lambda$ model (see Table~\ref{t:models}) for the evolution  between
those two redshifts. We expect this to be a good approximation for all
the models we test here because  the redshift difference between the
decoupling and the drag epoch is relatively small, and the sound
horizon at decoupling and drag is mostly governed by the fractional
difference between the number of photons and baryons.

\begin{figure}
\includegraphics[width=84mm]{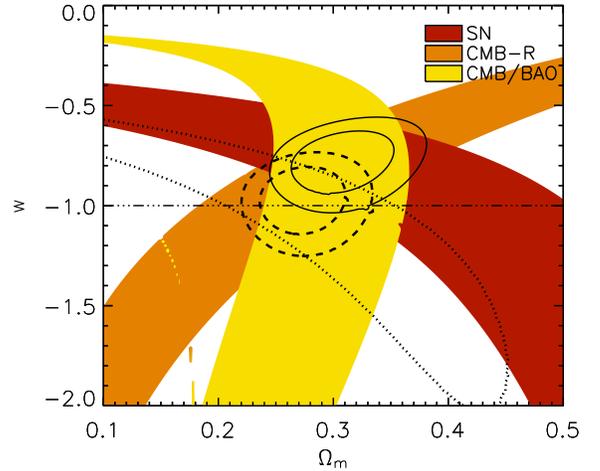}
\caption{
Flat dark-energy model (F$w$): a flat universe with constant
$w$. The constraint from each of the observational probes is shown by
shaded contours.  These are all 95\% confidence intervals for two
parameters.  Overlaid with black lines (95\% and 99.9\% confidence
intervals) are contours from combining CMB/BAO-{$\ell_A$}, CMB-${\cal R}$ and SN 
constraints.
The shaded contour labeled SN is for the analysis using the MLCS 
light-curve fitter. 
In this plot we have also added the CMB-$\cal R$ constraints,
although these are not included in the model selection.  
The dotted supernova contours are using the SALT-II fits. 
For the SALT-II dataset the combined contours are given by the dashed contours,
and are clearly in better agreement with the cosmological-constant value, 
$w=-1$, shown by the dashed-dotted line. 
}
\label{fig:smfcomb}
\end{figure}

\subsection{Combining the Datasets}

To clarify how we combine the data, we show in Figure~\ref{fig:smfcomb} our 
best fit to the data in the $w-\om$ plane  
for a flat universe with constant $w$ 
(The F$w$ model, see Table~\ref{t:models}). 
The constraints from each of the observational probes are shown 
by contours (according to the figure legend). 
In this and in all similar figures, 
these are $95\%$ confidence intervals for two parameters. 
The supernova dataset fitted with MLCS is shown by the shaded (red) contours, 
while the SALT-II SN analysis is shown by the dotted contours.
The combined contours ($95\%$ and $99.9\%$ confidence intervals) 
are overlaid in black for MLCS and by the dashed contours for SALT-II.
In our calculations, we are marginalizing over a 
common magnitude shift for all SNe, 
thus allowing for arbitrary values of the Hubble
constant, $H_0$, and the typical absolute magnitude of SNe~Ia.

This standard cosmology case is the only one for which we
also show the CMB-$\cal R$ constraint. 
The constraints from the
prescription for the BAO that we have implemented, which also 
include the CMB-$\ell_A$ parameter, are labeled CMB/BAO in the figures.
Using this CMB/BAO product cancels out some of 
the dependence on the sound horizon
size at last scattering. This thereby 
removes the dependence on much of the
complex pre-recombination physics that is needed to determine 
that horizon scale.

In Figure~\ref{fig:smfcomb}, the likelihood analysis takes 
into account the correlation between
$\ell_A$ and ${\cal R}$. In other words, our likelihood plot shows
contours derived from CMB-${\cal R}$ separate from contours derived
from BAO/CMB-$\ell_A$ that are then combined taking the weak correlation
into account. For the other models, and for the model selection, 
the CMB-${\cal R}$ is not included.
It is then the combination of
BAO/CMB-$\ell_A$ with SN data that we use as the basis for the model
comparison performed in this paper.   
This new analysis is
more model independent, and therefore better suited for ranking exotic
models. Excluding the $\cal R$ constraint also leaves more room for models
with non-zero curvature.

It was noted by
\citet[][]{percival07} 
that the distance
between the two BAO redshift bins seemed to be in slight conflict with
the supernova measurements for the standard model.  
Compared to the Percival et al. (2007) BAO/CMB constraints
this tension was
exacerbated by our new MLCS SN dataset.
The new BAO data ease some of this tension \citep[see also][]{percival09,hubert09}.
Rather than selecting only one of the BAO redshift measurements,
or any one of the SN light-curve fitters for that matter,
we choose to explore the implications of such tension for the more 
exotic models of interest in this paper.

Figure~\ref{fig:smfcomb} 
also helps us understand in a more quantitative way how the 
different datasets constrain the cosmology. 
The combined fit using MLCS for the SNe 
(the SALT-II case will be discussed further 
in Section~\ref{sect:salt}) and also CMB-$\cal R$ gives 
$w=-0.79\pm0.13$, 
while the more 
general
method we employ for all models in this paper 
(omitting $\cal R$) gives a best fit at 
$w=-0.83\pm0.24$.
The error bars quoted are 95$\%$ confidence level for one parameter, and we
only consider statistical errors.
For the approach used by K09 (SN+CMB-$\cal R$+BAO-$A$) 
we find 
$w=-0.78\pm0.13$. 
Although the procedure we have chosen to combine the different observational
constraints also affects the results,
it is clear that it is the MLCS analysis of the supernova dataset that 
makes the main difference in pushing $w$
away from the cosmological-constant value of $w = -1$.
We therefore move on to explore more exotic cosmological alternatives.

\begin{deluxetable}{lll}
\tablewidth{0pc}
\tablecolumns{4}
\tablecaption{Summary of models}
\tablehead{
\colhead{Model}  &
\colhead{Abbrev.} &
\colhead{Parameters\tablenotemark{a}} 
}
\startdata
Flat cosmo. const.  & F$\Lambda$ & $\om$  \\
Cosmological const. & $\Lambda$  & $\om$, $\ok$    \\
Flat constant $w$   & F$w$       & $\om$, $w$   \\
Constant $w$        & $w$        & $\om$, $\ok$, $w$      \\
Flat $w(a)$         & F$wa$      & $\om$, $w_0$, $w_a$     \\
Cardassian          & Ca         & $\om$, $q$, $n$     \\
Flat Chaplygin      & FCh        & $A$   \\
Chaplygin           & Ch         & $\ok$, $A$     \\
Flat Gen. Chaplygin & FGCh       & $A$, $\alpha$  \\
Gen. Chaplygin      & GCh        & $\ok$, $A$, $\alpha$     \\
DGP                 & DGP        & $\ok$, $\orc$ \\
Flat DGP            & FDGP       &  $\orc$ \\
LTB Gauss           & LTBg & $\Omega_{\rm in}$, $\Omega_{\rm out}$, $r_{0}$ \\
LTB Sharp           & LTBs & $\Omega_{\rm in}$, $\Omega_{\rm out}$, $r_{0}$
\enddata
\tablenotetext{a}{
The free parameters in each model. 
When fitting the SN Ia data we also fit an additional parameter, 
$\cal{M}$, for the normalization of SN magnitudes.  
We include this in the number of degrees of freedom and in 
the number of free parameters considered when calculating 
information criteria, but do not list it here as a parameter in each model.
For more details of these models and the parameters, see D07.
}
\label{t:models}
\end{deluxetable}

\section{Testing Non-standard Models}\label{sect:specific}

\cite{kessler09} introduced the first-year SDSS-II SNe 
to supplement existing data in order to constrain the standard 
cosmological model. 
They concentrated on testing a cosmological-constant model and 
a flat universe model with constant equation-of-state parameter.
Here we will extend that analysis to a more general
investigation  including several of the most popular non-standard
models. The models are briefly presented below.

\subsection{Beyond-Einstein Models}

There are many specific models based on new fundamental
physics that make predictions for the expansion history of the
universe.  We follow D07 in using information criteria to 
rank these models; see also the recent investigations by,
e.g., \cite{kurek08} and \cite{rubin08}.

With the new SDSS-II supernova data and updated data on the 
first-peak position of the CMB, combined with the new 
approach to the BAO constraints, we show below that it is
worthwhile retesting the non-standard cosmological models examined by
\citet{davis07}. 

The selected models
include both exotic dark-energy models as well as alternative models
that can be interpreted in terms of modifying the theory of gravity. 
The first class of models we test are the standard
cosmological constant, constant $w$, and variable
$w$ models. We also include Dvali-Gabadadze-Porrati (DGP), 
Cardassian expansion, and several versions of Chaplygin gas models. 
The appropriate references and the equations we use for
describing $H(z)$ in each of these models were collected by \citet[][
Equations 7--18]{davis07}, and we refer the reader to that paper for
further information. For reference, in Table~\ref{t:models} we list 
all the different models included in this study.\\

\subsection{The Lema\^{i}tre-Tolman-Bondi Models}

In addition to the above models, we also test some inhomogeneous cosmologies. 
Such models have gathered significant interest in recent
years as a means to explain the cosmological observations without
invoking dark energy. 
In the simplest class of such models we live close to the center of a
large, spherically symmetric void described by the
Lema\^{i}tre-Tolman-Bondi (LTB) metric 
\citep{lemaitre33,lemaitre97,tolman34,bondi47}.
The apparent acceleration of the expansion is then caused by
the spatial gradients in the metric, such that our local region has a
larger Hubble parameter than the outer region. While the LTB models challenge
the Copernican principle, several studies have shown that they
cannot be ruled out by present observational constraints
\cite[e.g.,][]{alnes06,enqvist07,troels08,caldwell08}.

The LTB models are characterized by two arbitrary functions, often
expressed as the expansion rate $H(r,t)$ and the matter density
parameter $\Omega_{\rm M}(r,t)$, which depend 
not only on time, but also on the radial coordinate. 
As a consequence, inhomogeneities arise independently in the matter 
distribution and in the expansion rate.

We will consider LTB models constrained by the requirement that
the Big Bang occurred simultaneously throughout space 
by implementing a particular choice of~$H_{0}(r)$,
\begin{equation} 
H_{0}(r)=\frac{3H_{0}}{2}\Bigg[ \frac{1}{\Omega_{
k}(r)}-\frac{\Omega_{\rm M}(r)}{\sqrt{\Omega_{k}^{3}(r)}}\sinh
^{-1}{\sqrt{\frac{\Omega_{k}(r)}{\Omega_{\rm M}(r)}}} \Bigg]\ ,
\end{equation} so that the time of the Big Bang was
$t_{BB}=\frac{2}{3}{H_{0}}{^{-1}}$ 
for all
observers irrespective of their position in space. 
The model is then
completely specified by only one free function, the matter density
parameter $\Omega_{\rm M}(r)$, with $\Omega_{\rm M}(r)+\Omega_{k}(r)=1$. 
We consider two different density profiles. In the first
model, $\Omega_{\rm M}(r)$ takes the form of a Gaussian underdensity,

\begin{equation} \Omega_{\rm M}(r)=\Omega_{\rm out}+(\Omega_{\rm
in}-\Omega_{\rm out})e^{-(r/r_{0})^{2}}\ .
\end{equation} 
This model has three free parameters, where $\Omega_{\rm
in}$ is the matter density at the center of the void, $\Omega_{\rm
out}$ is the asymptotic value of the matter density, and $r_{0}$ is the
scale size of the underdensity.

In the second model, $\Omega_{\rm M}(r)$ has a much sharper transition
from the local to the asymptotic value,
\begin{equation} \Omega_{\rm M}(r)=\Omega_{\rm out}+(\Omega_{\rm
in}-\Omega_{\rm out})\Bigg( \frac{1+e^{-r_{0}/\Delta
r}}{1+e^{(r-r_{0})/\Delta r}} \Bigg)\ .
\end{equation} 
Here $r_{0}$ characterizes the size at which the
transition occurs and the extra parameter $\Delta r$ characterizes the
transition width. In the limit where $\Delta r$ goes to zero, the
density profile becomes a step function. 
The two models are designated LTBg and LTBs in 
Table~\ref{t:models}.

Our sharp LTBs model is equivalent to the constrained model in
\cite{troels08}. We fix the transition
width to be $\Delta r = 0.065$ Gpc $h^{-1}$ to obtain a sharp transition, so
this is not a free parameter in our model.
Furthermore, in both our LTB models we allow $\Omega_{\rm
in}$ and $\Omega_{\rm out}$ to take any positive values $\le 1$,
i.e., the models need not be asymptotically flat and in principle we
allow also for solutions with a local \textit{overdensity}.

\section{Results}

\subsection{Model Ranking Using the MLCS Analysis}
\label{sect:mlcs}

The results of all our fits using the MLCS-fit SNe 
and the combined CMB/BAO constraints 
are summarized in Table~\ref{t:aic}. 
The corresponding results for the SALT-II-fit 
SNe are presented in Section~\ref{sect:salt}.
The results are stated in terms of $\chi^{2}$ and the 
given degrees of freedom, 
as goodness of fit (GoF), and in terms of
information-criteria (IC) assessments. 

The background for the use of IC 
for these models was reviewed by 
\citet[][ their Section 2]{davis07}. We follow their approach 
\citep[see also][]{liddle04} and use two IC
to select the best-fit models.  These are
the Bayesian information criterion (BIC)
\begin{equation} {\rm BIC} = -2 \ln{\cal L} + k \ln N, \end{equation}
and the Akaike information criterion (AIC)
\begin{equation} {\rm AIC} = -2 \ln {\cal L} + 2 k,\end{equation}
where ${\cal L}$ is the maximum likelihood,
$k$ is the number of parameters, and $N$ is the number of data points used in 
the fit.
A $\Delta$BIC larger than 6 would be considered unsupported as compared 
to the best model.

The number of degrees of freedom is derived from the 288 SNe and 
the two CMB/BAO measurements, less the contribution from $\cal{M}$ 
and the number of parameters listed in Table~\ref{t:models}.
Note that the fairly low $\chi^{2}$ per degree of freedom stems 
from the fact that we add an intrinsic dispersion to the supernova data.
As mentioned above, this is derived by measuring the
dispersion in the nearby sample.
The SDSS SNe actually have a lower dispersion (0.08 mag), 
which K09 attribute to selection effects (K09, their Appendix E).
We also ran a test with this lower intrinsic dispersion, 0.08 mag, and 
found that while the fits are now much worse (higher $\chi^{2}$ 
and very low GoF), 
the relative $\Delta$AIC are not affected by much. 
Some rank order differences would be seen in Table~2, but
only for differences of $\Delta$IC$ \lesssim$1, and these 
should not be regarded as significant.
Since the intrinsic dispersion has somewhat different meaning in the SALT 
and the MLCS frameworks, it makes little sense to compare the GoF for the 
different light-curve fitters.

\begin{figure}
\includegraphics[width=84mm]{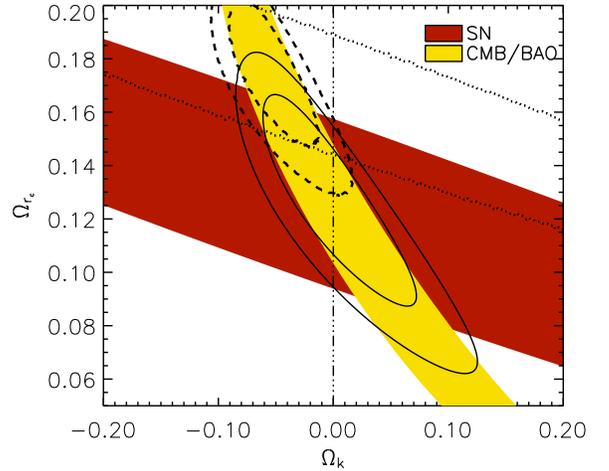}
\caption{
The constraints from SNe and CMB/BAO on the parameters in the DGP model.
The results have changed 
substantially from those of Davis~et~al.~(2007). 
This is both due to the new datasets 
and our choice of not using CMB-${\cal R}$.
The flat DGP model is indicated by the vertical dashed-dotted line; for the 
MLCS fit, it is the best-ranked model by the IC analysis.
The SALT-II fit to the SNe is again shown by the dotted contours.
The combined constraints using the SALT-II SNe outlined by the dashed contours
represent a poorer match to the CMB/BAO for the flat model.
}
\label{fig:dgpcomb}
\end{figure}

\begin{figure}
\includegraphics[width=84mm]{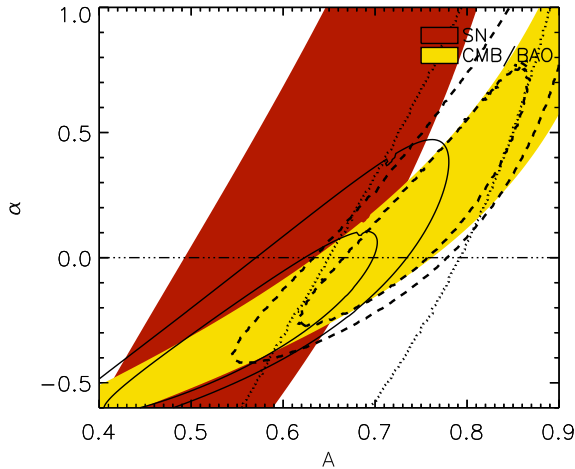}
\caption{
Constraints for the flat generalized Chaplygin gas (FGCh). 
From the constraints in this analysis, this model cannot be excluded. 
The dashed-dotted $\alpha=0$ line corresponds to 
parameters that match the $\Lambda$ model 
(with $\om=1-A$), and these do not match the combination with the MLCS fit 
data. 
The best MLCS fit is at 
$\alpha=-0.4$, $A=0.55$, which is far from the $\alpha=0.0$, $A=0.7$
that represents the best fit to the F$\Lambda$ model.
This is different from the analysis of D07 when the best fits were acquired 
for parameter values that mimic the cosmological constant. 
The combined constraints from the SALT-II SNe (dashed) are more consistent with
F$\Lambda$ values.
} 
\label{fig:chf}
\end{figure}

Compared to the analysis performed earlier by D07, 
we can see that the new MLCS dataset provides different
results in several interesting ways, as follows.

\begin{itemize}

\item
There are now a number of exotic models that fare rather well 
under the IC test. 
It is clear that both this dataset and the method we have used to combine the observational constraints 
permit a larger variety of models.
We note in particular 
that many of the models provide very similar $\chi$$^{2}$ values, and 
the IC tests therefore are sensitive to the number of parameters 
for the given model.\\

\item 
The simple, flat, cosmological-constant model (F$\Lambda$) 
favored by D07 is no longer on top of the ranking list in Table~\ref{t:aic}. 
This can of course be understood from Figure~\ref{fig:smfcomb}.
Instead, the model favored by both $\Delta$BIC and $\Delta$AIC 
is the flat DGP model that was unsupported by previous studies. 
In Figure~\ref{fig:dgpcomb} we can see the constraints on the DGP model
from the new data and analysis.  
These are very different from the case presented by D07.
The change is driven by the new SN data, and is exacerbated by the 
way we now combine the CMB/BAO and SN constraints 
(i.e. omitting CMB-${\cal R}$).

\item
The flat dark-energy model (F$w$) 
may still be a viable model for this new dataset in 
terms of $\Delta$IC.
However, in Figure~\ref{fig:smfcomb} 
we see that the most likely value from the combined dataset is 
$w = -0.83\pm0.24$ (95\% confidence level [C.L.]  for one parameter).
For this model, the value of $\om=0.31\pm0.06$.     
Including CMB-${\cal R}$ as in Figure~\ref{fig:smfcomb}, which is viable for this model, 
gives $w=-0.79\pm0.13$.
This offset from $w=-1$ is clearly a feature of the new MLCS analysis of 
the combined supernova dataset (see also K09).
The more general F$wa$ model gives
the best-fit value of 
$w_0=-0.73\pm 0.38$, 
while there are no useful constraints on the time varying component of
$w(a)$.

\item
Models such as the generalized Chaplygin gas were found by D07 to be good fits
to the data only when their parameters mimicked a cosmological constant.
The new best fits instead fall in regions of parameter space that do not
correspond to the cosmological constant. This is, for example, shown for the
flat generalized Chaplygin gas in Figure~\ref{fig:chf}.

The Cardassian expansion model is not very well 
constrained but 
also in this case the 
best fit does not
match the F$\Lambda$ model.
This agrees with the poorer rating of the cosmological-constant model 
in the present investigation.

\item
The only model that is very strongly unsupported 
compared to the other models is the flat Chaplygin~gas. 

\end{itemize}

\begin{figure}
\begin{center}
\includegraphics[angle=0,width=.48\textwidth]{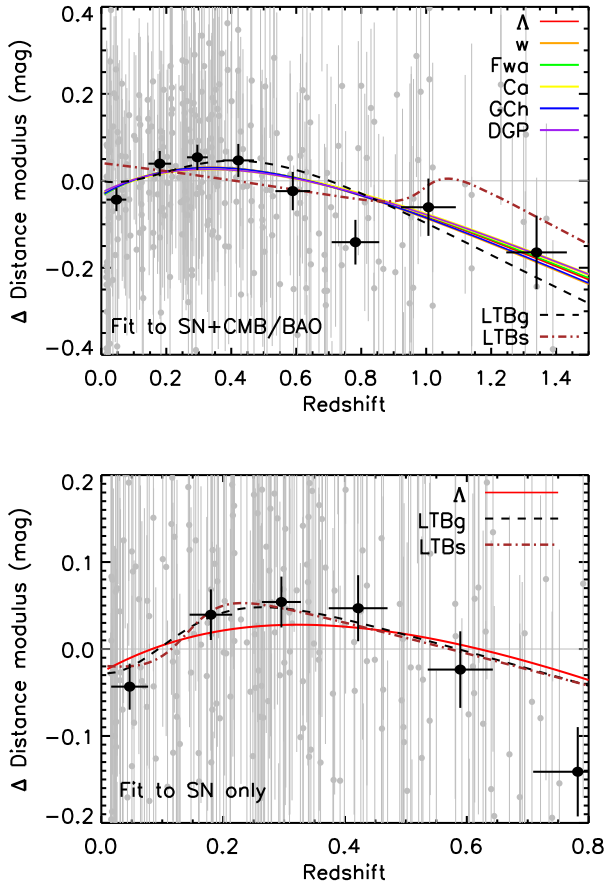}
\caption{  
(Upper) 
Hubble diagram for the MLCS supernova analysis. Distance
modulus differences in magnitudes 
are shown with respect to an empty universe. The grey points
are the SN~Ia distance moduli with error bars, while the black
points are binned data. 
The binning is done using $n$$\Delta$$z$=constant,
where $n$ is the number of SNe in the bin and $\Delta$$z$ the
redshift range. 
This uniform and unbiased binning was introduced by \citet{riess07}
to enable visualizing the Hubble diagram with bins distributed over the entire
redshift range, and with $n$$\Delta$$z=6$ we get 8 bins with error bars 
that are of similar size in both directions.
The redshift error bars show the standard deviation of
the redshifts in the bin, while the distance modulus error bars give
the standard deviation of the mean  of the distance modulus
uncertainties within the bin.
Note that the binning is only used for visualization in this figure, and that
all the calculations and fits are performed on the unbinned data.
We have also plotted the best fits for
a sample of models (we chose mainly the non-flat models in order not to
clutter the figure).  In the upper panel, these best fits are using both MLCS SNe and the
CMB/BAO constraints, i.e., they are not optimized for the SN Hubble
diagram alone. 
In order to plot these Hubble diagrams, 
the best fit value has been used for the different models. 
For example, for the upper panel and the $\Lambda$ model we derived 
$\cal{M}$-$M$ = $5\log{(c/H_0)}+25=43.35$. 
Here, $c$ is the speed of light, $H_{0}$ the Hubble constant 
(where $(c/H_0)$ is in~Mpc) and $M$ 
is the absolute magnitude of a Type Ia supernova.
The other models included in the figure have similar values 
for $\cal{M}$
($\sigma=0.02$~mag), 
with the LTBs model having the largest offset.
Note that in all the calculations, these nuisance parameters were 
marginalized over.
\\ 
(Lower) 
Hubble diagram for the best fits to the
MLCS SN~Ia data alone, relative to an empty universe. This panel is zoomed
in to focus on the sharp LTB (dotted)
and the Gaussian LTB (dot-dashed) models. 
For comparison we have also
plotted the best-fit $\Lambda$
model. 
}
\label{fig:hubblediagram}
\end{center}
\end{figure}

In addition to the contour plots, we provide in 
Figure~\ref{fig:hubblediagram} a Hubble diagram displaying a selection
of the discussed models. This representation offers 
perhaps a less abstract way
to judge the models. 
Note that the best-fit models are 
in the upper panel of Figure~\ref{fig:hubblediagram} 
constrained
both by the SNe~Ia and the CMB/BAO, i.e., the model parameters are not optimized
for a supernova Hubble diagram.

\begin{deluxetable}{lccccc}
\tablewidth{0pc}
\tablecolumns{5}
\tablecaption{Information-Criteria Results for MLCS}
\tablehead{
\colhead{Model}  &
\colhead{$\chi^2$/ dof } & 
\colhead{GoF (\%)} &
\colhead{$\Delta$AIC} &
\colhead{$\Delta$BIC} &
\colhead{$\Delta$$\Delta$AIC$_{Rv}$} 
}
\startdata
FDGP                &233.3   /    288 &       99.21   &    0.0   &    0.0  & $-2.3$ \\
$\Lambda$           &233.2   /    287 &       99.12   &    1.9   &    5.6  &  0.0   \\
DGP                 &233.2   /    287 &       99.12   &    2.0   &    5.6  &  0.4   \\
FGCh                &233.3   /    287 &       99.11   &    2.1   &    5.7  &  0.3   \\
Ch                  &233.4   /    287 &       99.10   &    2.1   &    5.8  & $-1.7$ \\
F$w$                &233.5   /    287 &       99.09   &    2.2   &    5.9  &  0.5   \\
F$wa$               &233.1   /    286 &       99.03   &    3.8   &    11.1 &  1.9   \\
 $w$                &233.2   /    286 &       99.02   &    3.9   &    11.2 &  1.0   \\
Ca                  &233.2   /    286 &       99.02   &    3.9   &    11.2 &  0.7   \\
GCh	            &233.2   /    286 &       99.01   &    3.9   &    11.3 &  0.9   \\
F$\Lambda$          &237.3   /    288 &       98.70   &    4.0   &    4.0  &  4.0   \\
LTBg                &235.5   /    286 &	      98.69   &    6.2   &    13.6 &  4.2 \\
LTBs                &237.6   /    286 &	      98.31   &    8.3   &    15.7 &  6.8 \\
FCh                 &257.6   /    288 &       90.09   &    24.3   &   24.3 &  13.2 
\enddata \tablecomments{ 
For our SN sample analyzed with MLCS plus CMB/BAO, the FDGP
model is preferred by both the AIC and the BIC.  The $\Delta$AIC and
$\Delta$BIC values for all other models in the table are then measured
with respect to these lowest values.  The goodness of fit (GoF)
approximates the probability of finding a worse fit to the data.  The
models are given in order of increasing $\Delta$AIC.  
The final $\Delta\Delta$AIC$_{Rv}$ column simply displays 
the difference in ranking when using different priors
in the MLCS supernova analysis, as described in Section~\ref{sect:extinction}. 
}
\label{t:aic}
\end{deluxetable}

\subsubsection{Changing the MLCS Priors}
\label{sect:extinction}

The MLCS light-curve fitter used by K09 is an 
updated version of the MLCS used by 
\citet[][ hereafter WV07]{WV07}. 
The new features include in particular new priors and an 
updated treatment of extinction.
As shown by K09, 
this new treatment significantly changes
the best-fit $w$ value for the F$w$ model, and this is not a feature 
of the new SDSS-II SN dataset.

To illustrate 
how these parameters within MLCS affect our cosmology ranking we have 
also performed 
calculations for an alternative set of assumptions. 
Following the discussion in K09 
(their Section 10.1.4 and Fig.~32) we start with the WV07 prior and then 
implement all of 
the cumulative changes up to (but not including) the point where K09 also 
change $R_V$ from the typical 
Milky Way value of 3.1 to the lower value of $R_V=2.2$. 
This is of course a rather arbitrary choice of parameter changes,
and is just meant to illustrate how different 
MLCS assumptions and implementations affect the fits to the 
Hubble diagram for the models tested in this paper. 
This particular choice corresponded to an offset in $w$ of $\lesssim0.1$ 
for the Nearby+ESSENCE+SNLS dataset analyzed using the  F$w$ model 
according to K09 (their Fig.~32). 
We perform here the calculations for all of our exotic 
models, for our choice of CMB/BAO prior, 
and for the complete SN dataset used in our analysis.

The results of the different assumptions in the MLCS are given
as the $\Delta\Delta$AIC$_{Rv}$ values in the last column of 
Table~\ref{t:aic}. 
This is simply the differences in $\Delta$AIC 
compared to our original MLCS analysis, i.e.,
$\Delta\Delta$AIC$_{Rv} = \Delta$AIC$_{\rm MLCS} - \Delta$AIC$_{\rm new\, priors}$, 
with both runs normalized to the best-fit 
model.
To not overload the presentation, we have chosen to include only the 
AIC statistics here, since this is an asymptotically unbiased estimator 
\citep{aic}.
The differences are 
not very large in this representation. 
Note again that differences of $\lesssim$1 in $\Delta$IC  are not significant.
All models fare slightly better compared to the FDGP model in this case.
A noticeable difference is that the
F$\Lambda$ model fares significantly 
better using this set of MLCS priors. 
In the F$w$ model, $w=-0.99\pm0.28$.  
We stress that we in no way prefer this set of 
parameters, but use them only to illustrate how such changes would 
affect our model selection.

\subsection{LTB}
\label{sect:ltbdiscussion}

Here we take 
the opportunity to  further discuss the results of our LTB models.
Combining the SN data with the CMB/BAO constraint provides a critical test of
the validity of our LTB models, since there appears to be some tension
between the best-fit parameter values preferred using the SNe and CMB/BAO data separately. 
This of course assumes that the BAO constraints can be used in this way also for 
LTB models \cite[see][]{troels08}.
The total $\chi^2$ values for our LTB fits are comparable to those of the
other models (Table~\ref{t:aic}), and because of the extra parameters the LTB
models fare poorly in the IC tests. 
These conclusions generally agree
with what \citet{troels08} found for their LTB models.

\begin{figure}
\includegraphics[width=84mm]{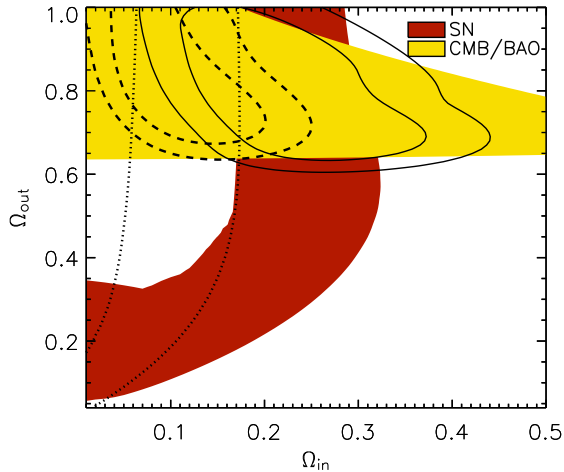}
\caption{The LTB Gauss model is able to give a reasonable fit to 
the data
but is not favored by the IC tests. It is clear that
the data prefer the matter density to vary from a low value locally to a
higher value in the distant universe. 
The scale size of the void in the best-fit model is 1.2 Gpc/$h$.
The SALT-II SNe (dotted) push $\Omega_{\rm in}$  to lower values while
leaving the constraint on $\Omega_{\rm out}$  unchanged.
}
\label{fig:ltb}
\end{figure}

Our best-fit models (using the MLCS supernova analysis) 
show the transition, $r_{0}$, 
at 1.2 (1.8) Gpc~$h^{-1}$ for the LTBg (LTBs) models. 
For the LTBg
the matter density $\Omega_{\rm in} =0.25$, $\Omega_{\rm out} =0.72$, 
and these numbers are similar for the LTBs model. This is illustrated in
Figure~\ref{fig:ltb}.

We note that some of the LTB models studied in the literature display  
distinct features in the low-$z$ end of the Hubble diagram
\citep[e.g.,][]{alexander07,clifton}.
It was therefore anticipated that the SDSS-II data would provide
decisive constraints by filling in the redshifts desert between 
$z\approx 0.1$ and $z\approx 0.4$.  
The smooth Hubble diagram in this redshift regime 
provided by our well-calibrated SDSS-II supernova dataset 
(see K09, their Fig.~23)
clearly disfavors 
such luminosity distance vs. redshift relations.

However, not all LTB models investigated possess such conspicuous
features \citep{alnes06,troels08}.  In fact, our simple LTB models can
fit the SN data (alone) with a $\chi^2$ somewhat lower than that of
the F$\Lambda$ model.  
This is illustrated in the Hubble diagrams in
Figure~\ref{fig:hubblediagram}, where
the LTB models are
included together with a selection of other alternative models. 
In the upper panel 
Hubble diagram all models have also been constrained  by the
CMB/BAO data. 
This combination produces a poorer fit for the LTB models and pushes 
the transition to higher redshifts.  
The lower panel instead displays the LTB Hubble
diagram when fit only to the supernova data, and demonstrates that a
good fit to the data can be obtained.

The more general conclusion of our LTB model test
is therefore that the SDSS-II SN set is useful for testing the LTB
models, but perhaps not decisive to the degree previously anticipated.  
Although the SDSS-II supernova set strongly constrains 
LTB models that display egregious features in the intermediate-redshift 
Hubble diagram, versions of the LTB model that do not display such 
features, such as the LTBg and LTBs models 
considered here, can be made to fit the supernova data.

However, in terms of our IC test, we have found that these 
LTB models are not well motivated; 
the number of extra parameters involved is not
justified by a better fit to the SN data, in particular not 
when the CMB/BAO constraints are also included. 
Note that these results also hold when using the SALT-II light-curve fitter to
analyze the SN data (Table~\ref{t:aicsalt}). 
We also mention that if we were to impose the prior
of asymptotic flatness on our LTB models, the $\chi^{2}$ for these fits
would increase even further.

\subsection{Model Ranking Using the SALT-II Analysis}\label{sect:salt}

We have also performed calculations using SN data fit with the completely
independent light-curve fitter SALT-II \citep{guy}, 
following the implementation as described by K09. 
K09 noted a substantial offset for the value of $w$
as calculated in the standard model using MLCS versus
SALT-II, 
and devote a long discussion to the differences of these light-curve fitters.
As seen in Figure~\ref{fig:smfcomb}, the SALT-II fits are 
more consistent with the concordance model, and we here
investigate in detail how such a difference alters our ranking of
the investigated exotic models.

In performing these calculations, 
we had to exclude four of the SNe that were well fit by MLCS
because in the SALT-II analysis they gave poor 
fits\footnote{The SNe designated d085, e020, k429, and HAWK.}.

The SALT-II prescription clearly makes the SN data more
compatible with the standard models. 
In Table~\ref{t:aicsalt} we see
that the highest-ranked model is the F$\Lambda$
model, a flat universe with a cosmological constant.  
It has a similar $\chi^{2}$ as many of the other models, 
and is primarily favored by the IC
since it has only one free parameter.
This echoes the
conclusions of D07. In the rightmost column of Table~\ref{t:aicsalt} 
we provide the
difference in $\Delta$AIC compared to the MLCS results, to illustrate
which models benefit most from using the SALT-II SN set. 
Note that the numbers of degrees of freedom are less than for the MLCS 
table both since we had to exclude four SNe, and due to the two extra free 
parameters in the SALT-II fits. 
We note the following:

\begin{itemize}
\item
F$\Lambda$ evidently fares much better with SALT-II than with MLCS, 
as was already evident from Figure~\ref{fig:smfcomb}.  
For the F$w$ model the best-fit value for the
equation-of-state parameter is 
$w=-1.14\pm0.26$ (95$\%$ C.L. for one parameter), 
with $\Omega_{\rm m}=0.30\pm0.05$.
Including CMB-${\cal R}$ as in Figure~\ref{fig:smfcomb}, which is viable for this model, 
gives $w=-0.98\pm0.14$.

\item
On the contrary, the FDGP model is now pushed toward the bottom of
the ranking list.
Figure~\ref{fig:dgpcomb} again illustrates the situation, with
the SALT-II SN constraint dragging the solution along the CMB/BAO
constraint and away from the flat universe. The non-flat DGP model is
still one of the models that fares better in this investigation than
in D07 (as is the Ch model). 

\item
The LTB models fare as poorly (or worse) compared to 
the best-ranked models when the SALT-II SN data are used. 
In this respect the conclusions from Section~\ref{sect:ltbdiscussion} 
hold independent of the adopted light-curve fitter.

\item
The SN distances 
obtained using the SALT-II light-curve fitter provide results that
are overall more consistent with the  analysis of D07.
For the FGCh model (Figure~\ref{fig:chf}), we can see
that the SALT-II SN distances allow 
a match to the CMB/BAO constraints for parameter values
that are more consistent with those expected for the 
cosmological-constant model.
\end{itemize}

\section{Systematics}

When performing the same kind of IC ranking 
of cosmological models as in D07, the new
datasets and methods explored in this paper lead us to
quite different conclusions.
When we use the MLCS analysis 
the standard cosmological-constant
model no longer stands out as the best and simplest model, 
and new room is made for a range of more exotic models.
The results from the SALT-II analysis are quite different.

A basic underlying assumption in the analysis 
is that
the errors are mainly statistical. 
The systematic errors in, for example, the SN surveys are extensively
discussed elsewhere \citep[e.g.,][]{astier06,WV07,kessler09}, and the
uncertainties include, in particular, the distribution of host-galaxy
extinctions ($A_V$) and the dust-reddening properties ($R_V$).

While model selection techniques
are not well equipped to face systematic errors, we can regard the 
above-mentioned calculations as a 
simplistic perturbation theory approach to test how assumptions 
regarding light-curve fitters affect the model selection. 

In Table~\ref{t:aic} we illustrate how the model selection is affected by 
rather arbitrarily changing the priors in the MLCS light-curve fitter. 
Given the change in results engendered by using the ratio of 
total-to-selective absorption of $R_V = 2.2$ that K09 derived from the SN sample
itself, such a low value of $R_V$ 
should clearly find support also from independent
astrophysical observations.

The most dramatic
differences noted in this investigation come from 
the very different results obtained for the
two light-curve fitters investigated. This clearly points to unresolved
systematic differences that need to be further explored.
Table~\ref{t:aicsalt} shows in the final column that the fits of 
some exotic models are dramatically affected by the choice of light-curve
fitter in the analysis of the SN data.

While we acknowledge the way \citet[][]{kowalski08} 
treat the SN data from their large Union08
compilation, we also note that adding a dispersion to the
error mimics the use of more free parameters in the fit. 
Using the same Union08 dataset, \citet[][]{rubin08} reached conclusions
similar to the ones in this paper; the difference in $\chi^{2}$ values for
different cosmological models are rather small. 
We have not included any systematic errors in our tests, and 
including larger systematic errors would clearly make it even more difficult to
differentiate between cosmological models for a given dataset.

However, it is also clear from our investigation 
that trying to encapsulate the systematic errors 
is notoriously difficult. The substantial disagreement
in model selection obtained with different light-curve fitters 
investigated here demonstrates that
deeper understanding of the systematics is warranted.
It is also important to 
refrain from simply choosing the method or datasets that 
confirm previous findings or present prejudices.\\

\begin{deluxetable}{lccccc}
\tablewidth{0pc}
\tablecolumns{6}
\tablecaption{Information-Criteria Results for SALT-II}
\tablehead{
\colhead{Model}  &
\colhead{$\chi^2$/ dof } & 
\colhead{GoF (\%)} &
\colhead{$\Delta$AIC} &
\colhead{$\Delta$BIC} &
\colhead{$\Delta$$\Delta$AIC$_{\rm MLCS}$} 
}
\startdata
F$\Lambda$ &  274.4 /      282 &      61.63   &    0.0  &     0.0  &  4.0    \\
FGCh       &  273.0 /      281 &      62.32   &    0.6  &     4.3  &  1.5    \\
$\Lambda$  &  273.4 /      281 &      61.66   &    1.0  &     4.6  &  0.9    \\
DGP        &  273.4 /      281 &      61.65   &    1.0  &     4.7  &  1.0    \\
Ch	   &  273.9 /      281 &      60.87   &    1.5  &     5.1  &  0.7    \\
F$w$       &  274.0 /      281 &      60.67   &    1.6  &     5.3  &  0.6    \\
GCh        &  273.0 /      280 &      60.61   &    2.6  &     10.0 &  1.3    \\
F$wa$	   &  273.4 /      280 &      60.05   &    3.0  &     10.3 &  0.8    \\
Ca         &  273.4 /      280 &      59.97   &    3.0  &     10.3 &  0.9    \\
$w$	   &  273.5 /      280 &      59.82   &    3.1  &     10.4 &  0.8    \\
LTBg       &  276.5 /      280 &      54.71   &    6.2  &     13.5 &  0.0    \\
FCh	   &  281.9 /      282 &      49.06   &    7.5  &     7.5  &  16.8   \\
FDGP       &  282.5 /      282 &      48.05   &    8.1  &     8.1  & $-8.1$  \\
LTBs       &  289.3 /      280 &      33.90   &    18.9 &     26.2 & $-10.6$ 
\enddata 
\tablecomments{ 
For our SN sample analyzed with SALT-II plus CMB/BAO, the 
F$\Lambda$ model is preferred by both the AIC and the BIC.   
The $\Delta$AIC and $\Delta$BIC values for all other models in the table
are then measured with  respect to these lowest values.  The goodness
of fit (GoF) approximates the probability of finding a worse fit to
the data.   The models are given in order of increasing $\Delta$AIC.
The numbers of degrees of freedom are less than for the MLCS fits, since
four SNe were omitted and because SALT-II uses two extra fit parameters.
The final column simply displays the AIC differences 
as compared to the MLCS results in Table~\ref{t:aic}, i.e.,
$\Delta\Delta$AIC$_{\rm{MLCS}}$=$\Delta$AIC$_{\rm{MLCS}}$$-$$\Delta$AIC$_{\rm{SALT}}$.
}
\label{t:aicsalt}
\end{deluxetable}

\section{General Expansion History}\label{generalised}

Given the difficulty in assessing particular models, we also consider
some ways to derive more general cosmological results from our
dataset. The usual parameter-fitting techniques used in cosmology are
limited by the necessity of assuming a model with parameters to
fit;  there is more information in the cosmological data than can be
extracted by parameter fitting. In this section we will ignore
specific dark energy theories and instead try to elucidate directly the 
expansion history of the universe, $H(z)$, or the evolution of an
equation-of-state parameter of dark energy, $w(z)$. The SN dataset fitted
with MLCS is used here, since this easily provides distance moduli
without having to fit a specified cosmology for the $H(z)$ section,
and it avoids additional free parameters for the $w(z)$ analysis.

\subsection{The Equation-of-State Parameter $w(z)$}\label{w(z)}

We start this investigation by assuming a piecewise constant
equation-of-state parameter, in order to fit a more general
dark-energy model. The value $w_i$ is calculated in each redshift bin
$i$ (defined by the upper limit $z_i$), where we have 
arbitrarily chosen $z_i = [0.15,0.6,1.7,1090]$, roughly
corresponding to low-, mid-, high-$z$ SNe and CMB data,
respectively.  We have fitted both $w_i$ and $\om$
using a Monte Carlo Markov Chain, assuming a flat universe.

Note that the $w_i$ are correlated; the covariance matrix is not diagonal.
We can, however, decorrelate the $w_i$ estimates by 
\citep[following][]{huterer05}
changing the basis through an orthogonal matrix rotation that
diagonalizes the covariance matrix. This corresponds to applying a
weighting function to the $w_i$. The uncorrelated $w_i$ are thus
linear combinations of all $w_i$ described by the weight function.

That is, appropriately weighted data from all redshifts
are used to obtain $w$ in each bin, and 
in our case, the value of $w_i$ corresponding to low redshifts 
($w_1$) is a linear combination of the value of 
$w(z)$ at $z<0.15$
($\sim82\%$ contribution), $w(0.15<z<0.6)$ ($\sim18\%$ contribution),
and negligible contributions from higher redshifts. 
In the same way,
$w_2$ is a 
combination of $w(z<0.15)$ ($\sim24\%$ contribution), $w(0.15<z<0.6)$
($\sim73\%$ contribution) and $w(0.6<z<1.7)$ ($\sim3\%$ contribution). 
The value of $w_3$ has a $\sim17\%$ contribution from
$w(0.15<z<0.6)$ and a $\sim87\%$ contribution from
$w(0.6<z<1.7)$, whereas $w_4$ predominantly comes from the highest redshift bin
and only has small negative contributions 
from the three lower redshift bins.

\begin{figure}
\includegraphics[width=84mm]{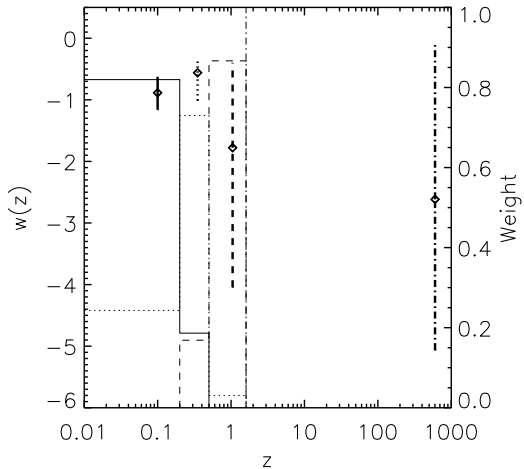} 
\caption{ 
Observational constraints on a piecewise
constant $w(z)$ plotted (diamonds) at the middle value in each redshift bin.
Note here that results are given not in terms of the original $w(z)$
but weighted with data from all redshift bins in order to
decorrelate the bins. The weights are displayed as thin lines
with styles corresponding to the respective points, and 
with the scale
given on the right-hand side of the figure. This is
further detailed in the text (Section~\ref{w(z)}).
The results include our CMB/BAO constraints. The error bars are
$68\%$. 
The constraints on the evolution of $w(z)$ at $z \gtrsim 1$ are
clearly very weak.
}
\label{fig:r3}
\end{figure}

The results are presented in Figure~\ref{fig:r3}, and are
discussed in Section~\ref{sect:generalinterp}.

\subsection{The Hubble Parameter $H(z)$}\label{h(z)}

\citet{wangtegmark05} proposed a technique to extract
the expansion history in uncorrelated redshift bins from SNe~Ia data. 
This technique was used to great effect by
\citet{riess07}.
We now apply it with the
new information from the SDSS-II supernova sample.

\begin{figure}
\includegraphics[width=84mm]{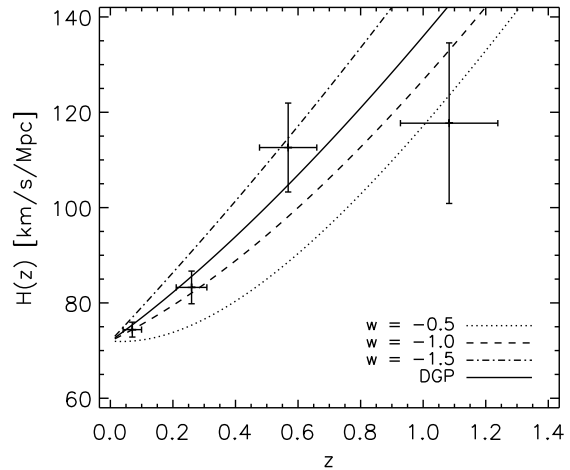}
\caption{
With the rich SN dataset we can also trace $H(z)$ 
in a more model-independent way. 
Here the $H(z)$ evolution is compared to some theoretical models. 
The models are normalized to agree at redshift zero, and
we have assumed $H_0 = 72$~km~s$^{-1}$Mpc$^{-1}$. 
The data points are normalized to have the lowest-redshift bin on the 
$w=-1$ curve, and the four different points are independent of each other.
Only supernova data were used for this plot, and a flat universe was assumed.
}
\label{fig:hz4bins}
\end{figure}

The results are displayed in Figure~\ref{fig:hz4bins}, 
where the theoretical models have been normalized as described in the caption.
The redshift bins
have been chosen by the requirement that the ratio of the error on the
Hubble parameter and the redshift range probed by each
measurement should be close to a constant times the redshift
derivative of the Hubble parameter; we want to keep a uniform
signal-to-noise ratio across the bins. Ideally, this constant should
be equal to one, corresponding to
\begin{equation}\label{eq:dhdz}
  \sigma_H = \frac{dH}{dz}\Delta_z\ ,
\end{equation}
where $\sigma_H$ is the uncertainty in the Hubble parameter and
$\Delta_z$ is the redshift range that each measurement probes,
as calculated by \citet{wangtegmark05}. We compute $dH/dz$ 
using a fiducial cosmological model, here taken to be a flat
universe with $\om =0.3$.

We find that when using three redshift bins, the error from the
width of the redshift bin dominates, whereas for four bins, the Hubble
parameter uncertainty is somewhat larger than the corresponding 
error from the redshift range probed.  Note that only the supernova
data are used in obtaining $H(z)$.

\begin{figure}
\includegraphics[width=84mm]{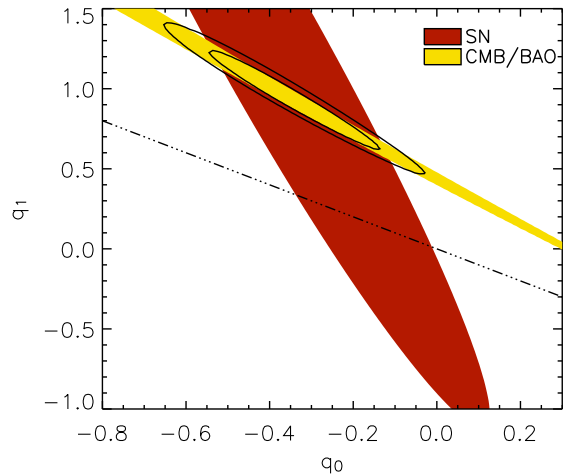} 
\caption{ 
Another less model-dependent investigation comes from
exploring the deceleration parameter and its evolution. Following
\cite{mortsellclark08}, we plot this representation together with our
new observational constraints. On the abscissa, the evidence
for current acceleration is evident.
The dashed-dotted line demarcates models with early deceleration (above the line)
from model with early acceleration; that the best-fit lies well above 
this line demonstrates evidence for past deceleration.  
}
\label{fig:q0qa}
\end{figure}

\subsection{The Deceleration Parameter $q$}\label{qoqa}

Another less theory-dependent investigation comes from
exploring the deceleration parameter and its evolution. 
Following \cite{mortsellclark08}, we consider a flat universe where
\begin{equation}\label{eq:=qparam}
q(a)=q_0+q_1(1-a)=q_0+q_1\frac{z}{1+z}\, .
\end{equation}
At zero
redshift, $q(a=1)=q_0$ and in the infinite past, $q(a=0)=q_0+q_1$.
This parametrization appears to be reasonably flexible
in the sense that performing a least-squares fit to the dark-energy 
models employed in this paper always provides an acceptable fit.
The comoving distance is given by
\begin{equation}
  d_c(z) = \int_{0}^{z}{\frac{dz}{H(z)}}=
\frac{1}{H_0}\int_{0}^{z}{\exp{\left[-\int_0^{v}\frac{[1+q(u)]du}{(1+u)}\right]}dv}\, ,
\end{equation}
and the luminosity distance, which is the relevant quantity for SN Ia
observations, is given by
\begin{equation}
  d_L(z) = \frac{1+z}{H_0\sqrt{|\ok|}}S_k{\left[ \sqrt{|\ok|} H_0 d_c(z)\right]} \, .
\end{equation}
The angular diameter distance, relevant for the scale of BAO and CMB,
is given by $d_A = d_L/(1+z)^2$.  We can now use Figure~\ref{fig:q0qa}
to study the expansion history of the universe within this simple flat
model, as discussed below. 

\subsection{General Results}\label{sect:generalinterp} 

The method of presenting
the evolution of $w(z)$ and $H(z)$ given in
Figs.~\ref{fig:r3}  and \ref{fig:hz4bins} may have more general and
long-lasting value than fits to specific models.

The $H(z)$ evolution shown in Figure~\ref{fig:hz4bins} is displayed together
with a few cosmological models.
We can see how the DGP model produces an acceptable fit to
the evolution of the Hubble parameter. The general trend, an increasing value
for $H(z)$ with increasing redshift, is clear.

Any $w(z)$ evolution would be an obvious signature of
exotic models. 
We do not, however, see any statistically significant 
evolution of $w$ in 
Figure~\ref{fig:r3} \citep[see also][]{kowalski08}. 
This representation could also be
used to constrain generic ``freezing'' and ``thawing'' models, following, 
e.g., \citet[][]{sullivan07,caldwell07}, 
and in this respect Figure~\ref{fig:r3} can be seen as an illustration
of our ability to model-independently 
constrain the evolution of $w$ using current data. 
We note that the result that $w$ is consistent with $-1$ across the
redshift range does not depend on the choice of binning, although the
size of the error bars does.
Using Eq. 11 we derive for our
combined SNe~Ia and CMB/BAO constraints 
(Figure~\ref{fig:q0qa}) 
$q_0=-0.34\pm0.16$ and $q_1=0.93\pm0.25$.  
The negative value of
$q_0$ demonstrates that the present universe is accelerating. Note,
however, that with the current (MLCS) dataset, this conclusion does
not follow for this parametrization for the SN dataset
alone. Furthermore, the fact that the combined contour clearly hovers
above the dashed-dotted line demonstrates that the universe was also
decelerating in the past. The redshift where the universe
transits from a decelerated expansion to an accelerated expansion is
constrained to 
$0.41 < z_t < 0.72$ (95$\%$ C.L. for one parameter).

\begin{figure}
\includegraphics[width=84mm]{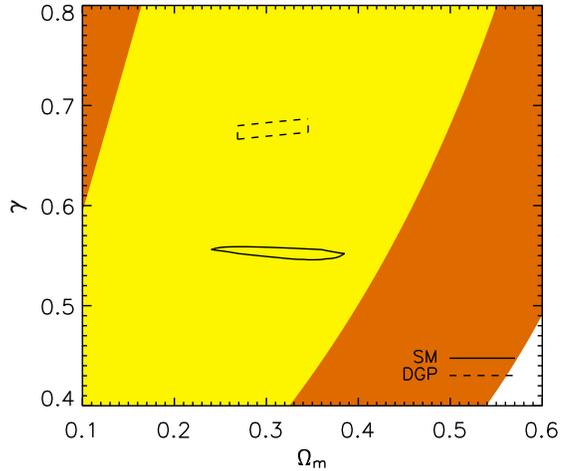}
\caption{
The growth of structure is an independent way 
to constrain cosmological models. 
This is illustrated here for standard gravity (SM) and for the DGP model
(see Section~\ref{gamma}).
It is clear that the present data do not yet select between these models, 
but any future deviations from $\gamma=0.55$ 
above a few percent would indicate deviations from ordinary gravity.
Solid and dashed lines are 95\% confidence levels for one parameter while 
the filled yellow (light) and orange (dark) regions show 68.3\% and 95\%
confidence levels for two parameters respectively.
}
\label{fig:gengrav}
\end{figure}

\subsection{Growth Factor, $\gamma$}\label{gamma}

There are a number of more exotic models that cannot be ruled out,
even in principle, if the magnitude-redshift evolution we observe
remains consistent with $\Lambda$CDM.  This is because they can mimic
standard dark energy perfectly in this respect. 

For many of these models, it would be useful to investigate additional
and complementary constraints that can distinguish them from
$\Lambda$CDM.  This can be done using information about structure
formation as expressed by the growth factor \citep{linder05,
linder07}. It is a measurable parameter of a model, in the same sense
as the matter density $\om$ or the equation-of-state parameter $w$. 
All models in which general relativity holds have a growth parameter 
$\gamma \approx 0.55$, 
but models that explain the acceleration by modifying gravity
may have a significantly different growth factor. Thus, this approach 
has the
potential to separate models that have identical predictions for
expansion history but differ in the form of gravity.

We show in Figure~\ref{fig:gengrav} the predicted growth factor for
the DGP model and for the standard model (SM=F$w$, dashed and solid contours, respectively), 
together with constraints
from current data (filled contours).  
The DGP model has the same number of parameters as
the cosmological-constant model, and can therefore provide a useful
framework for testing modified gravity alternatives. The DGP model
fared very well in the IC test in Table~2, but does give different
predictions for the growth of structure. 

Figure~\ref{fig:gengrav} uses the new data from \citet{guzzo08} analyzed 
following 
the procedure of \citet{linder07}.
The filled contours are growth-factor data ($f\propto \Omega_m^\gamma$) 
using $f=0.49\pm0.14$ at $z=0.15$ and 
      $f=0.91\pm0.36$ at $z=0.77$. 
It is clear that the present data do not yet select between these models, 
but any deviations from $\gamma=0.55$ 
above a few percent would indicate departures from ordinary gravity.

\section{Conclusions}

We have followed the information-criteria model selection outlined in
D07 using the new supernova dataset from the SDSS-II (K09) 
and a new way to combine the updated BAO and CMB datasets. 
The ranking using the MLCS supernova analysis is quite different from the 
models selected by D07, with more room for exotic models. 
Using instead the SALT-II light-curve fitter, we obtained fits more
consistent with the cosmological constant.  
Some LTB models were also tested; 
they were able to give 
decent fits to the SN data, but
were not favored by the full IC tests.

We also discuss how more general results can be derived from this new
dataset, and demonstrate that the expansion of the universe is indeed speeding up now,
while it used to be slowing down. This conclusion is based
on combining constraints from the supernova data with complementary
data.

The information-criteria tests used in this analysis are fairly simple, but
serve the purpose to quantitatively rank models with rather similar 
$\chi^{2}$-fit values. 
The way the results differ between the different light-curve 
fitters suggests that more detailed statistical tests are not justified. 

The data from the releases of several major supernova searches
(SDSS-II, ESSENCE, SNLS, {\it HST}) have been analyzed in this paper. Most of
these teams plan to release substantial updates of their supernova
samples soon. This will be important in improving the
statistics, but our analysis also makes clear that systematic
effects, in particular uncertainties in the treatment of extinction
and light-curve fitting, will limit the predictive power in selecting 
a model based on these data. Rather than determining $w$ with
increased accuracy, sometimes claimed to be measured to within
$\sim6\%$ \cite[e.g.,][]{kowalski08,komatsu08}, this work 
cautions on the importance of the systematic effects. 

We note that the 
recent investigation by \citet{hicken09}
also finds 
differences in the derived value for $w$ using
different light-curve fitters and also for different cuts in the
datasets, and provide several good suggestions on how to improve
future supernova cosmology in this regard.  The SDSS-II dataset is
indeed of sufficiently high quality to enable further studies of the
relevant systematics.

We have explored how systematics limit our ability to select the best
models for the universal expansion based on current data.  
A discussion on the benefits of MLCS versus SALT-II is beyond the scope
of this paper. K09 trace part of the discrepancies to the way the
light-curve fitters treat extinction, how this affects very blue SNe
and in particular to uncertainties in the bluest pass-bands.  Better
constraints on the systematics from the final data releases from 
SDSS-II, ESSENCE, and SNLS may well help in resolving some of the tension
discussed in this paper.  We may then find that our standard model
holds.  Or we may simply live in a more exotic universe.

\section*{Acknowledgments}
DARK is funded by the Danish National Research Foundation. 
The Oskar Klein Centre is funded by The Swedish 
Research Council. J.S. and E.M. acknowledge support by the 
Swedish Research Council. 
A.V.F. is grateful for the support of NSF grants
AST-0607485 and AST-0908886.  
J.S. is a Royal Swedish Academy of Sciences Research Fellow 
supported by a grant from the Knut and Alice Wallenberg Foundation.
We wish to thank Rick Kessler for valuable discussions and contributions to this paper.
We also thank 
Will Percival 
for important information on the BAO constraints,
and Giorgos Leloudas for comments.
We acknowledge use of the easyLTB code generously provided by
J. Garcia-Bellido and T. Haugb{\o}lle.
    
Funding for the creation and distribution of the SDSS and SDSS-II
has been provided by the Alfred P. Sloan Foundation,
the Participating Institutions,
the National Science Foundation,
the U.S. Department of Energy,
the National Aeronautics and Space Administration,
the Japanese Monbukagakusho,
the Max Planck Society, and the Higher Education Funding Council for England.
The SDSS Web site \hbox{is {\tt http://www.sdss.org/}.}

The SDSS is managed by the Astrophysical Research Consortium
for the Participating Institutions.  The Participating Institutions are
the American Museum of Natural History,
Astrophysical Institute Potsdam,
University of Basel,
Cambridge University,
Case Western Reserve University,
University of Chicago,
Drexel University,
Fermilab,
the Institute for Advanced Study,
the Japan Participation Group,
Johns Hopkins University,
the Joint Institute for Nuclear Astrophysics,
the Kavli Institute for Particle Astrophysics and Cosmology,
the Korean Scientist Group,
the Chinese Academy of Sciences (LAMOST),
Los Alamos National Laboratory,
the Max-Planck-Institute for Astronomy (MPA),
the Max-Planck-Institute for Astrophysics (MPiA), 
New Mexico State University, 
Ohio State University,
University of Pittsburgh,
University of Portsmouth,
Princeton University,
the United States Naval Observatory,
and the University of Washington.

This work is based in part on observations made at the 
following telescopes.
The Hobby-Eberly Telescope (HET) is a joint project of the University of Texas
at Austin,
the Pennsylvania State University,  Stanford University,
Ludwig-Maximillians-Universit\"at M\"unchen, and Georg-August-Universit\"at
G\"ottingen.  The HET is named in honor of its principal benefactors,
William P. Hobby and Robert E. Eberly.  The Marcario Low-Resolution
Spectrograph is named for Mike Marcario of High Lonesome Optics, who
fabricated several optical elements 
for the instrument but died before its completion;
it is a joint project of the Hobby-Eberly Telescope partnership and the
Instituto de Astronom\'{\i}a de la Universidad Nacional Aut\'onoma de M\'exico.
The Apache 
Point Observatory 3.5 m telescope is owned and operated by 
the Astrophysical Research Consortium. We thank the observatory 
director, Suzanne Hawley, and site manager, Bruce Gillespie, for 
their support of this project.
The Subaru Telescope is operated by the National 
Astronomical Observatory of Japan. The William Herschel 
Telescope is operated by the 
Isaac Newton Group on the island of La Palma
in the Spanish Observatorio del Roque 
de los Muchachos of the Instituto de Astrofisica de 
Canarias. The W.M. Keck Observatory is operated as a scientific partnership 
among the California Institute of Technology, the University of 
California, and the National Aeronautics and Space Administration. The 
Observatory was made possible by the generous financial support of the 
W. M. Keck Foundation.

\end{document}